\documentclass[12pt]{iopart}
\usepackage{graphicx,amssymb}

\newcommand{\LL}{\mathcal{L}}
\newcommand{\HH}{\mathcal{H}}

\begin{document}

\title{Constraints on Gauss-Bonnet Gravity in Dark Energy Cosmologies}

\author{Luca Amendola$^1$, Christos Charmousis$^2$ and Stephen C Davis$^{3,4}$}

\address{${}^1$ INAF/Osservatorio Astronomico di Roma, 
Viale Frascati 33, 00040 Monte Porzio Catone (Roma), Italy}

\address{${}^2$ LPT, Universit\'{e} Paris--Sud, B\^{a}t.\ 210, 
91405 Orsay CEDEX, France}

\address{${}^3$ ITP, \'{E}cole Polytechnique F\'{e}d\'{e}rale de
Lausanne, CH--1015 Lausanne, Switzerland}
\address{${}^4$ Lorentz Institute, Postbus 9506, NL-2300 RA Leiden, 
The Netherlands}

\eads{\mailto{amendola@mporzio.astro.it},
\mailto{Christos.Charmousis@th.u-psud.fr} and
\mailto{sdavis@lorentz.leidenuniv.nl} }

\begin{abstract}
Models with a scalar field coupled to the Gauss-Bonnet Lagrangian
appear naturally from Kaluza-Klein compactifications of pure higher-dimensional
gravity. We study linear, cosmological  perturbations in the limits
of weak coupling and slow-roll, and derive simple expressions for
the main observable sub-horizon quantities: the anisotropic stress
factor, the time-dependent gravitational constant, and the matter perturbation
growth factor. Using present observational data, and assuming
slow-roll for the dark energy field, we find that the fraction of
energy density associated with the coupled Gauss-Bonnet term cannot
exceed 15\%. The bound should be treated with caution, as there are
significant uncertainies in the data used to obtain it. Even so, it
indicates that the future prospects for constraining the coupled Gauss-Bonnet
term with cosmological observations are encouraging. 

\end{abstract}

\maketitle

\section{Introduction}

In recent years there has been a renewal of interest in scenarios
that propose alternatives or corrections to Einstein's four dimensional
gravity. These proposals are of differing origin as well as motivation:
some are based on multidimensional theories in which gravity propagates
in more than four dimensions (see for example~\cite{arka}), others
on scalar-tensor couplings, that could give cosmologically observable
effects~\cite{wet88}, or violate the equivalence principle~\cite{dam}.
Another class of theories modify gravity by adding terms in the Lagrangian
that depend on quadratic combinations of the Riemann tensor (see for
example~\cite{gb}), such as $R^{2}$ and $R_{\mu\nu}R^{\mu\nu}$,
yielding higher than second-order field equations. Among the applications
of such a generalisation we notice that these terms have been used
in the past to model inflation~\cite{star} and dark energy~\cite{rn}.
As it is well-known, there is a unique quadratic combination of the
Riemann curvature tensor that, if added to the usual Einstein-Hilbert
action, does not increase the differential order of the equations
of motion~\cite{lovelock,zumino}. For historical reasons this is
called the Gauss-Bonnet (GB) term as its geometric origin can be traced
back to the Gauss-Bonnet theorem regarding the Euler characteristic
of two dimensional surfaces~\cite{spivak}. 

In classical gravity theories involving dimensions higher than
four there is no good reason to omit it (apart from complication).
Studies of higher dimensional
gravity, mostly in the context of braneworlds, have shown that such
theories have many surprising properties. Without trying to be exhaustive,
we can refer to black hole physics~(e.g.~\cite{bh}), gravitational
instabilities~(e.g.~\cite{neg}), and the dynamics of co-dimension
one and higher braneworlds~(e.g.~\cite{gbr}).

In four dimensions the Gauss-Bonnet combination reduces to a total
divergence and, as such, is dynamically irrelevant. However, in the
case of a scalar-tensor theory (see for example~\cite{DF} for a
detailed and general discussion), the Gauss-Bonnet term couples to
the scalar sector and four dimensional gravity is modified (see
also~\cite{kal} for the case of Lorentz and Chern-Simons terms). A coupled
Gauss-Bonnet term should therefore be included in the most general
second-order scalar-tensor theory. Such a coupling is manifest if
we consider, for example, the case of when the scalar field is a modulus
field originating from a Kaluza-Klein (KK) toroidal (i.e.\ flat)
compactification of a $4+N$ dimensional theory. This is explicitly shown in
section~\ref{sec:model}. There it is also shown that in order to be
consistent with the KK truncation, higher-order scalar field terms
must also be included; a fact that is very often neglected in the literature.

The above discussion about uniqueness of the Gauss-Bonnet term 
is, strictly speaking, valid only at the classical level. 
However, additional motivation
stems from low-energy effective string actions, such as
the one for heterotic string theory. At tree level (for string coupling) 
this includes the Gauss-Bonnet term as the leading order 
and unique (up to that order) ghost-free $\alpha'$ correction to the 
Einstein-Hilbert term~\cite{gross} (see for example~\cite{bin} for
cosmology and braneworld models emanating from such actions).

Since the coupled Gauss-Bonnet term is to be naturally included in
a scalar-tensor theory, an important question arises: how can current
observations constrain such models? It has already been shown in the
literature that at the background level a coupled Gauss-Bonnet term
allows for a viable cosmology~\cite{sa}. In this article we address
the question of the evolution of linear cosmological perturbations,
and see whether interesting constraints can be derived with present
or future data. Another, and maybe more stringent way to set up
constraints is to look for modifications of gravity on local scales;
parameterized post-Newtonian (PPN) analysis~\cite{gdot,Will}
requires a next to leading order study of slow-moving sources around
the Earth's gravitational field, or of solar system bodies. Such a
study would necessarily be dependent on the details of the Gauss-Bonnet
coupling, since it is dimensionful, and as we will see would be
independent of the cosmological expansion. In fact in this case the
relevant background would be that of flat spacetime, or of a modified
Schwarzschild metric. This study is beyond the scope of the
present paper (see~\cite{Gilles} for a first approach in this
direction) and here we will stick to cosmological constraints which
are as model free as possible.

As a first and most conservative approach, and in order to disentangle
the Gauss-Bonnet effects from the pure scalar sector of the theory,
we place ourselves in the Einstein frame%
\footnote{We will assume however that the energy-momentum tensor is conserved
i.e., that the Einstein frame is the physical frame.%
}. Since we assume the scalar field is driving an accelerated expansion,
we also endow the field with a sufficiently flat potential in a cosmological
background.

In  section~\ref{sec:model} we describe the class of theories
we will be using, and give the cosmological field equations in suitable
notation. We also derive the form of the theory when the scalar field
is of Kaluza-Klein origin. In section~\ref{sec:Gs}, which is the
backbone of this work, we write the cosmological perturbation equations.
We then analyse the field equations for small Gauss-Bonnet coupling,
and for a slowly rolling scalar field. Present-day constraints are discussed
in section~\ref{sec:obs}. We conclude and discuss our results
in section~\ref{sec:conc}.

\section{Second-order scalar-tensor theory}

\label{sec:model}

Let us consider the following Lagrangian \begin{equation}
\LL=\frac{1}{16\pi G}\left(R-[\nabla\phi]^{2}-2U(\phi)+\alpha\LL^{(c)}\right)+\LL_{\mathrm{mat}}\label{eq:L}\end{equation}
 where \begin{equation}
\LL^{(c)}=\xi_{1}(\phi)\LL_{GB}
+\xi_{2}(\phi)G_{\mu\nu}\nabla^{\mu}\phi\nabla^{\nu}\phi
+\xi_{3}(\phi)[\nabla\phi]^{2}\nabla^{2}\phi
+\xi_{4}(\phi)[\nabla\phi]^{4}
\label{melina}\end{equation}
 \begin{equation}
\LL_{GB}=R^{2}-4R_{\mu\nu}R^{\mu\nu}
+R_{\alpha\beta\mu\nu}R^{\alpha\beta\mu\nu}\, .\end{equation}
$G_{\mu\nu}$ is the Einstein tensor and $\LL_{GB}$ is the Gauss-Bonnet term. 
The coupling constant $\alpha$ has dimension $\mathrm{mass}^{-2}$. Note that we
place ourselves in the Einstein frame. Under this hypothesis the above
expression~(\ref{eq:L}) is the most general second-order ghost-free
scalar-tensor Lagrangian (with uncoupled matter sector). In this article
we will be interested in the observational effects coming purely from
higher-order gravity, and so we will take
$\delta\LL_{\mathrm{mat}}/\delta\phi=0$. 
This hypothesis, along with the fact that we place ourselves in the
Einstein frame, ensures that all deviations from standard gravity will
originate from the $\LL^{(c)}$ sector in (\ref{melina}).

The form of the functions $\xi_{i}$ will be related to the origin
of the scalar field. One possibility is the compactification of extra
dimensions, as was mentioned in the introduction. Consider the general
$4+N$ dimensional Lagrangian of pure gravity, 
\begin{equation}
\LL^{4+N}\propto\left(R-2\Lambda+\alpha\LL_{GB}\right)
\label{hendrix}\end{equation}
which is of second order in the curvature operator%
\footnote{For $N>2$ to keep all generality~\cite{lovelock} we would have
to add higher-order curvature invariants. To understand their geometric
origin see~\cite{zumino}.}
 and yields second-order field equations, which are divergence free
and have well defined and stable perturbations around the vacuum.
We use the metric anzatz, 
\begin{equation}
ds^{2}=e^{\mu\phi(x)}g_{\mu\nu}(x)dx^{\mu}dx^{\nu}
+dX_{a}dX^{a}e^{-2\mu\phi(x)/N}\, ,
\label{marion}\end{equation}
where $\mu^{2}=2N/(2+N)$ and $dX_{a}dX^{a}$ is the Euclidean
metric, and also locally that of the $N$-torus. Compactifying the
$N$ extra dimensions then gives the above theory~(\ref{eq:L}),
with $\xi_{1}^{-1}\propto U\propto e^{\mu\phi}$, and 
\begin{equation}
\xi_{2}=\frac{8}{2+N}\xi_{1}\, , \qquad
\xi_{3}=\frac{2\sqrt{2}(2-N)}{\sqrt{N(2+N)}}\xi_{1}\, , \qquad
\xi_{4}=\frac{4(1-N)}{(2+N)N}\xi_{1}\, .
\label{xiKK}\end{equation}
 The scalar field $\phi$ plays the role of the overall size of the
$N$-torus as we can see from (\ref{marion}). We see that typically
all of the four possible ghost-free quadratic-order gravity terms will be
present, and not just the Gauss-Bonnet term as is often assumed (see
also~\cite{daviz}).

The general field equations have been given in~\cite{cartier}. Here we
write down the equations for a Friedmann-Robertson-Walker (FRW)
metric, including pressureless matter. We set the conformal Hubble
function $\mathcal{H}=aH(t)$ and adopt the $e$-folding time 
$\eta=\log a$ as time variable:  
\begin{equation} \fl 
3\HH^{2}=8\pi G\rho a^{2}+Ua^{2}+\frac{1}{2}\HH^{2}\dot{\phi}^{2}-\frac{\alpha\HH^{4}\dot{\phi}}{2a^{2}}\left[24\xi_{1}'-9\xi_{2}\dot{\phi}-6\xi_{3}\dot{\phi}^{2}-(3\xi_{4}-\xi_{3}')\dot{\phi}^{3}\right]\label{eq:Fried}\end{equation}
\begin{eqnarray} \fl
\frac{\dot{\HH}}{\HH}+1  =  8\pi G\frac{\rho a^{2}}{6\HH^{2}}+2\frac{Ua^{2}}{3\HH^{2}}-\frac{1}{6}\dot{\phi}^{2}-\frac{\alpha\HH\dot{\HH}\dot{\phi}}{2a^{2}}\left[12\xi_{1}'-3\xi_{2}\dot{\phi}-\xi_{3}\dot{\phi}^{2}\right]\nonumber \\
    {}-\frac{\alpha\HH^{2}\dot{\phi}^{2}}{6a^{2}}\left[12\xi_{1}''-3\xi_{2}'\dot{\phi}-\xi_{3}'\dot{\phi}^{2}\right]-\frac{\alpha\HH^{2}\ddot{\phi}}{2a^{2}}\left[4\xi_{1}'-2\xi_{2}\dot{\phi}-\xi_{3}\dot{\phi}^{2}\right]\label{eq:acc1}\end{eqnarray}
\begin{eqnarray} \fl
\ddot{\phi}+\dot{\phi}\left(2+\frac{\dot{\HH}}{\HH}\right)  =
-\frac{a^{2}U'}{\HH^{2}}+\frac{\alpha\HH\dot{\HH}}{a^{2}}\left[12\xi_{1}'-9\xi_{2}\dot{\phi}-9\xi_{3}\dot{\phi}^{2}-2(3\xi_{4}-\xi_{3}')\dot{\phi}^{3}\right]
\nonumber \\ \hspace{-0.4in}
  {}-\frac{\alpha\HH^{2}\dot{\phi}^{2}}{2a^{2}}\left[3\xi_{2}'+4\xi_{3}'\dot{\phi}+(3\xi_{4}'-\xi_{3}'')\dot{\phi}^{2}\right]-\frac{\alpha\HH^{2}}{a^{2}}\ddot{\phi}\left[3\xi_{2}+6\xi_{3}\dot{\phi}+2(3\xi_{4}-\xi_{3}')\dot{\phi}^{2}\right]\label{eq:kg0}\end{eqnarray}
 (dots denote $d/d\eta$, primes denote $d/d\phi$). We now introduce
the dimensionless variables
\begin{eqnarray} \fl
\Omega_{m}=\frac{8\pi G\rho a^{2}}{3\mathcal{H}^{2}}\, ,\qquad
\Omega_{K}=\frac{\dot{\phi}^{2}}{6}\, , \qquad
\Omega_{P}=\frac{Ua^{2}}{3\HH^{2}}\, , \qquad
\Omega_{1}=-\frac{4\HH^{2}}{a^{2}}\alpha\dot{\phi}\xi_{1}'\, ,\qquad
\nonumber \\
\Omega_{2}=\frac{3\HH^{2}}{2a^{2}}\alpha\dot{\phi}^{2}\xi_{2}\, , \qquad
\Omega_{3}=\frac{\HH^{2}}{a^{2}}\alpha\dot{\phi}^{3}\xi_{3}\, ,\qquad
\Omega_{4}=\frac{\HH^{2}}{6a^{2}}\alpha\dot{\phi}^{4}(3\xi_{4}-\xi_{3}')
\, ,
\end{eqnarray}
which are subject to the Friedmann equation (\ref{eq:Fried}), so

\begin{equation}
\sum\Omega_{i}=1 \, .
\end{equation}
Defining $\Omega_{\phi}=1-\Omega_{m}$,
equation~(\ref{eq:acc1}) gives the acceleration of the universe to be
$\dot{\HH}/\HH=-[3\Omega_{\phi}w_{\phi}+1]/2$ where 
\begin{eqnarray} \fl
\Omega_{\phi}w_{\phi}\left(1-\frac{3\Omega_{1}+2\Omega_{2}
+\Omega_{3}}{2}\right) = -\Omega_{P}+\Omega_{K}
\nonumber \\ \hspace{-0.6in} {} 
-\left(\frac{\dot{\phi}\xi''_{1}}{\xi'_{1}}+\frac{\ddot{\phi}}{\dot{\phi}}-\frac{5}{2}\right)\frac{\Omega_{1}}{3}-\left(\frac{\dot{\phi}\xi'_{2}}{2\xi_{2}}+\frac{\ddot{\phi}}{\dot{\phi}}-\frac{3}{2}\right)\frac{4\Omega_{2}}{9}-\left(\frac{\dot{\phi}\xi'_{3}}{3\xi_{3}}+\frac{\ddot{\phi}}{\dot{\phi}}-\frac{3}{2}\right)\frac{\Omega_{3}}{3}+\frac{\Omega_{4}}{3}
 \label{eq:weff}\end{eqnarray}
 and $w_{\phi}$ gives the effective equation of state for all the
$\phi$-dependent terms, including the quadratic-order gravity ones.
It is clear that the density fractions $\Omega_{i}$ quantify
the deviation from standard gravity and standard cosmology. 
It is important to notice that since  the $\xi_i$ couple to $\HH$, the
densities $\Omega_i$ will disappear from the field equations 
whenever $\HH$ is zero; in other words for $a(t)=\mathrm{constant}$. This
signifies that we need a FRW background which is neither Minkowski nor
de-Sitter in order for the higher-order terms to affect the
cosmological evolution. This spells out why one needs to go beyond
linear order in order to find local effects of the Gauss-Bonnet term,
at least for a flat spacetime. With this in mind the goal
of this article will be to express observable quantities in terms
of the cosmological $\Omega_{i}$.

\section{Cosmological perturbation equations and the linear post-Newtonian
limit}

\label{sec:Gs}

We derive in this section the cosmological perturbation equations;
more precisely, we derive the limit for a weak gravitational field
(i.e.\ linearised perturbation equations), with slow time-variation
(we specify below how slow), and at small scales (with respect to
the horizon's size). We refer to this as the linear post-Newtonian
(LPN) limit. It is convenient to use the longitudinal gauge metric
written directly in $\eta=\log a$:
\begin{equation}
ds^{2}=e^{2\eta}\left[-(1+2\Psi)\frac{d\eta^{2}}{\HH^{2}}
+(1-2\Phi)dx_{i}dx^{i}\right]\, .
\end{equation}
The field $\Psi$ is the Newtonian potential and $\Phi$ is the leading-order, 
spatial post-Newtonian correction which will permit the calculation
of the stress-anisotropy. We always assume that the effective mass
of the field $\phi$ is very small, which is what is expected for
a dark energy field; the precise requirement is that the length scale
$1/m_{\phi}$ be much larger than the typical scales of the experiment
used to set the constraints.

In practice, the LPN limit amounts to considering 
$\triangle\Phi\gg\mathcal{H}^{2}\times(\Phi,\dot{\Phi},\ddot{\Phi})$
and similarly for the other gradient terms. For a plane wave perturbation
of wavelength $\lambda$ we see that $H^{2}\Phi$ is much smaller
than $\triangle\Phi$ when $\lambda\ll1/H$. The requirement that
$\dot{\Phi}$ is also negligible implies the condition
 \begin{equation}
\frac{d\log\Phi}{d\log a}\ll(\lambda H)^{-2} \, ,
\end{equation}
which clearly holds in perturbation theory for any reasonable perturbation
growth as soon as $\lambda\ll1/H$ (notice for instance that the gravitational
potential $\Phi$ is in fact constant in a standard matter dominated
universe, or in a spherical collapse, and slowly varying in most cosmological
models, e.g.\ those with a cosmological constant). The same arguments apply
for $\ddot{\Phi}$ as well as the other metric potential $\Psi$, and
also for $\delta\phi$. The $(\eta\eta)$ and $(ij)$ components
of the gravitational field equations, and the scalar field equation,
can be written in the following compact fashion, \begin{equation}
{\frac{3}{2}}\HH^{2}\delta\Omega_{m}=A\,\triangle\Phi+B\,{\frac{\triangle\delta\phi}{\dot{\phi}}}\label{eq:000gr}\end{equation}
 \begin{equation}
A\,\triangle\Psi=C\,\triangle\Phi+D\,{\frac{\triangle\delta\phi}{\dot{\phi}}}\label{eq:ij0gr}\end{equation}
 \begin{equation}
B\,\triangle\Psi=D\,\triangle\Phi
-E\,{\frac{\triangle\delta\phi}{\dot{\phi}}} \, .
\label{eq:kggr0}\end{equation}
The background coefficients are given by 
\begin{eqnarray}
A=1-\Omega_{1}-\frac{\Omega_{2}}{3}\, ,\qquad 
B={\frac{1}{2}}\left(\Omega_{1}+\frac{4\Omega_{2}}{3}
+\Omega_{3}\right) 
\, , \nonumber \\
C=1+\Omega_{1}\left(1-\frac{\dot{\HH}}{\HH}-\frac{\ddot{\phi}}{\dot{\phi}}-\dot{\phi}{\frac{\xi_{1}''}{\xi_{1}'}}\right)+\frac{\Omega_{2}}{3}
\, , \nonumber \\
D=\Omega_{1}\frac{\dot{\HH}}{\HH}+\frac{2\Omega_{2}}{3}\left(\frac{\dot{\HH}}{\HH}+\frac{\ddot{\phi}}{\dot{\phi}}+\frac{\dot{\phi}}{2}{\frac{\xi_{2}'}{\xi_{2}}}\right)
\, , \nonumber \\
E={\frac{1}{2}}\left(\dot{\phi}^{2}
+\frac{2\Omega_{2}}{3}\left[1+2\frac{\dot{\HH}}{\HH}\right]
+2\Omega_{3}\left[1+\frac{\dot{\HH}}{\HH}
+\frac{\ddot{\phi}}{\dot{\phi}}
+\frac{\dot{\phi}}{3}{\frac{\xi_{3}'}{\xi_{3}}}\right]+4\Omega_{4}\right)
\, , \nonumber \\
\hspace{-0.3in} F={\frac{1}{2}}\left(\dot{\phi}^{2}+
\left[1-\frac{\xi''_{1}\dot{\phi}}{\xi'_{1}}\right]\Omega_{1}
+2\left[1-\frac{\xi'_{2}\dot{\phi}}{3\xi_{2}}\right]\Omega_{2}
+3\left[1-\frac{\xi'_{3}\dot{\phi}}{9\xi_{3}}\right]\Omega_{3}
+4\Omega_{4}\right)
\, .
\end{eqnarray}
 For completeness we also include the $(0j)$ component of the gravitation
field equations in the same limit 
\begin{equation}
\frac{3}{2}\HH^{2}\theta\Omega_{m}=(B-A)\,\triangle\Psi
+F\frac{\triangle\delta\phi}{\dot{\phi}}-A\triangle\dot{\Phi}
-B\frac{\triangle\dot{\delta\phi}}{\dot{\phi}}\, .
\end{equation}
 Finally, the energy-momentum conservation equations are as usual
$\dot{\delta}=-\theta$ and $\dot{\theta}+\theta+(\dot{\HH}/\HH)\theta=-\triangle\Psi/\HH^{2}$,
where $\delta\equiv\delta\rho/\rho$ is the matter density contrast
and $\theta$ is the divergence of the matter peculiar velocity field.

In Einstein gravity, the perturbation equations in the LPN limit would
give the usual expression of the Poisson equation (\ref{eq:000gr}) and
the condition $\Psi=\Phi$, i.e.\ the vanishing of the stress
anisotropy $\sigma=\Psi-\Phi$ (\ref{eq:ij0gr}).
To quantify the deviation from Einstein perturbations we will express
our results by two observable quantities: the time variation of
Newton's constant, $\dot{G}/G$, and the fractional anisotropic stress
$\gamma=\Phi/\Psi$. Making use of the perturbation equations we find
\begin{equation}
\gamma={\frac{EA+BD}{{D^{2}+CE}}}\, .\end{equation}
Poisson's equation is \begin{equation}
\triangle\Psi=\frac{G_{*}(\eta)}{G}\frac{3}{2}\HH^{2}\delta\Omega_{m}
\, , \label{eq:Poisson}\end{equation}
where the variable Newton's {}``constant'' is
\begin{equation}
\frac{G_{*}(\eta)}{G}=\frac{D^{2}+CE}{A^{2}E+2ABD-B^{2}C}\, .
\end{equation}
 From Poisson's equation, we see that the growth of the matter fluctuations
is immediately given by the equation
\begin{equation}
\ddot{\delta}+\left(1+\frac{\dot{\HH}}{\HH}\right)\dot{\delta}
-\frac{3}{2}\frac{G_{*}(\eta)}{G}\delta\Omega_{m}=0
\end{equation}
In the absence of any quadratic-order gravity terms ($\alpha=0$) one has
$\gamma=1$ and $\delta\phi=0$, returning a constant effective 
gravitational coupling ($G_{*}=G$). Notice in passing that the corrections to
gravity cancel in the limit $\HH\to0$, which shows that the
contribution from the coupled Gauss-Bonnet term vanishes at the
linear level for this limit. 

Actual experiments set up to constrain the variation of $G$ (and the
value of $\gamma$) are generically {\it local experiments} set-up in
the immediate vicinity of the local gravitational field of the  Earth
and the Sun~\cite{gdot,Will}. Also cosmological constraints on $\dot{G}$
from primordial nucleosynthesis constrain the variation in the
expansion rate induced by a variable $G$ in an otherwise standard
Friedman equation (see e.g.~\cite{villante}). In the present case,
however, the Friedman equation does not contain a variable $G$ since
we did not include the Brans-Dicke term. Had we done so it would be
difficult to disentangle the effect of the scalar field from the
Gauss-Bonnet correction. In the actual state of experiment it turns
out that the only way to observe our modified $G$ is to constrain the
form of the Poisson equation from cosmological observations.

The Poisson equation of the GB theory is similar to the one in Brans-Dicke
gravity which (in the Einstein frame) arises from the coupling of
the scalar field to matter or (in the Jordan frame) from scalar corrections
to gravity as a generalised Lagrangian $\mathcal{L}(\phi,R)$. It
is clear that the specific form and time dependence of the correction
to the Poisson equation $G_{*}(\eta)$, depends on the form of the
Lagrangian. We expect $\gamma-1$ and $\dot{G}_{*}$ to be small when
$\Omega_{i}$ are small, and so we will now examine several
different limiting situations for which this is the case, either as
a result of small $\alpha$, or slow-roll (small $\dot{\phi}$).

\subsection{Small-$\alpha$ limit}

Let us begin with the small-$\alpha$ limit. If we include terms which
are of linear order in $\alpha$, we find that $\delta\phi$ and $\sigma$
are no longer zero. Solving the perturbation
equations (\ref{eq:ij0gr}) and (\ref{eq:kggr0}),
we find that in this limit the stress anisotropy parameter is \begin{equation}
\gamma\approx\frac{A}{C}=1-\Omega_{1}\left(2-\frac{\dot{\HH}}{\HH}-\frac{\ddot{\phi}}{\dot{\phi}}-\dot{\phi}{\frac{\xi_{1}''}{\xi_{1}'}}\right)-\frac{2\Omega_{2}}{3}+\Or(\alpha^{2})\, .\label{eq:ac}\end{equation}
 In general, the constraint on the Gauss-Bonnet densities $\Omega_{i}$
will depend on the specific dynamics. Similarly, the gravitational
constant is \begin{equation}
\frac{G_{*}(\eta)}{G}\approx\frac{C}{A^{2}}=1+\Omega_{1}\left(3-\frac{\dot{\HH}}{\HH}-\frac{\ddot{\phi}}{\dot{\phi}}-\dot{\phi}{\frac{\xi_{1}''}{\xi_{1}'}}\right)+\Omega_{2}+\Or(\alpha^{2})\, .\label{eq:Gsa}\end{equation}
 There are also in principle $\Or(\alpha\triangle\delta\phi)$ corrections
to the expression for $G_{*}$, although we see from the scalar field
equation~(\ref{eq:kggr0}) that $\triangle\delta\phi=\Or(\alpha)$,
and so its contribution to $G_{*}$ is of quadratic order in $\alpha$.
The higher-order gravity terms $\Omega_{3}$ and $\Omega_{4}$ only
affect $G_{*}$ via $\delta\phi$, and so similarly their contributions
are quadratic order in $\alpha$. This is despite the fact that their
contributions to the Friedmann equation are of the same order as $\Omega_{1}$.

Differentiating the expression~(\ref{eq:Gsa}), and using the field
equations (in the small $\alpha$ limit) to simplify it, we find 
\begin{eqnarray} \fl
\frac{\dot{G}_{*}}{G_{*}}=\left(3\Omega_{P}\frac{U'}{U}
\left[\frac{U''}{U'}+3\frac{\xi_{1}''}{\xi_{1}'}\right]+10\frac{\dot{\HH}}{\HH}
-10+5\frac{\ddot{\phi}}{\dot{\phi}}+11\dot{\phi}\frac{\xi_{1}''}{\xi_{1}'}
-\dot{\phi}^{2}\frac{\xi_{1}'''}{\xi_{1}'}\right)\Omega_{1}
\nonumber \\ {}
+\left(\dot{\phi}\frac{\xi_{2}'}{\xi_{2}}+2\frac{\dot{\HH}}{\HH}
-2+2\frac{\ddot{\phi}}{\dot{\phi}}\right)\Omega_{2}+\Or(\alpha^{2})
\, . \label{dGsa}
\end{eqnarray}

Again, to derive a precise bound on $\Omega_{1,2}$ from this
expression requires the dynamics of $\phi$ to be specified,
i.e.\ a choice of potential and couplings. However, unless the couplings
and the potential are unnaturally steep functions of $\phi$, we can
assume that all the ratios $U''/U'$, $\xi_{1}''/\xi'_{1}$, and similar
are $\Or(1)$; then the expression simplifies to 
\begin{equation}
\frac{\dot{G}_{*}}{G_{*}}\approx\left(b_{1}+b_{2}\dot{\phi}+
b_{3}\dot{\phi}^{2}+5\frac{\ddot{\phi}}{\dot{\phi}}\right)\Omega_{1}
+\left(b_{4}+b_{5}\dot{\phi}+2\frac{\ddot{\phi}}{\dot{\phi}}\right)
\Omega_{2}+\Or(\alpha^{2})\, ,
\label{dGsaa}\end{equation}
where all the $b_{i}$'s are quantities of order unity.

We see that the coefficients of $\Omega_{1,2}$ are smallest in
absolute value in the slow-roll limit, when
$\ddot{\phi}/\dot{\phi},\dot{\phi}\ll1$ 
(assuming there is no {}``chance'' cancellation of the coefficients).
With this in mind, we will now turn to cosmologies where the only
assumption is slow-roll for $\phi$.

\subsection{Slow-roll limit}

An assumption which is justified in the context of dark energy models
is to suppose that the scalar field is in a slow-rolling regime. This
allows us to specify to some extent the background dynamics of the
field. Slow-roll is realised for $\ddot{\phi}/\dot{\phi},\dot{\phi}\ll1$.
Under this hypothesis we can solve the field equations keeping 
leading-order kinetic terms. Our aim is to generically obtain the role of
the higher-order terms of the Lagrangian, in the slow-rolling regime,
without any further assumptions. In particular we do not take $\alpha$
to be small, in contrast to the previous subsection.

The Friedmann equation truncated to linear order gives simply
\begin{equation}
1=\Omega_{m}+\Omega_{P}+\Omega_{1}+\Or(\dot{\phi}^{2})\, .
\label{miles}\end{equation}
In the same way the acceleration field equation~(\ref{eq:weff}) reduces to
\begin{equation}
\frac{\dot{\HH}}{\HH}=1-\frac{3}{2}\Omega_{m}-\frac{9\Omega_{m}+2}{4}\Omega_{1}+\Or(\dot{\phi}^{2})\label{bird}\end{equation}
 where we have used (\ref{miles}) to trade $\Omega_{P}$ for $\Omega_{m}$.
Note that a negative Gauss-Bonnet density $\Omega_{1}$ can assist
late time acceleration, and a positive one can suppress it. Substituting
(\ref{miles}) and (\ref{bird}) into the scalar field equation~(\ref{eq:kg0})
we obtain, 
\begin{eqnarray} \fl
\left[2\frac{\dot{\phi}U'}{U}(1-\Omega_{m})+(2-3\Omega_{m})\Omega_{1}\right]
\nonumber \\ \fl
{}+\left[(2-\Omega_{m})\dot{\phi}^{2}+2(2-3\Omega_{m})\Omega_{2}
-\frac{\dot{\phi}U'}{U}(1+\Omega_{m})\Omega_{1}
- 6\Omega_{m} \Omega_{1}^{2}\right]
+\Or(\dot{\phi}^{3},\ddot{\phi}/\dot{\phi})=0
\, .\label{keith}\end{eqnarray}
The leading-order part of the scalar field equation (\ref{keith}) constrains
the validity of the slow-roll limit. It is self consistent if, and only if,
\begin{equation}
\frac{U'}{U}=\frac{(3\Omega_{m}-2)}{2(1-\Omega_{m})}
\frac{\Omega_{1}}{\dot{\phi}}
=-\frac{2\alpha \xi'_1 \HH^{2}(3\Omega_{m}-2)}{a^{2}(1-\Omega_{m})} \, .
\end{equation}
This shows that for $\dot{\phi}\to0$ , $U'/U$ remains finite. Hence,
in contrast to conventional first-order gravity, it is possible to
have slow-roll even when $U$ is not particularly flat. We see that
to next order, (\ref{keith}) gives us $\Omega_{2}$ with respect
to $\Omega_{1}$ and the scalar field parameters,
\begin{equation}
4(1-\Omega_{m})\Omega_{2}=
-2\frac{(2-\Omega_{m})(1-\Omega_{m})}{2-3\Omega_{m}}\dot{\phi}^{2} 
+ \frac{13\Omega_{m}-2-9\Omega_{m}^{2}}{2-3\Omega_{m}}
 \Omega_{1}^{2} \, .
\label{eq:omega2}\end{equation}

Taking the LPN limit we obtain\begin{equation}
\gamma=1-\frac{\Omega_{1}^{2}(2-3\Omega_{m})(1-3\Omega_{m})}
{2\dot{\phi}^{2}+4\Omega_{2}(1-\Omega_{m})+\Omega_{1}^{2}(2-3\Omega_{m})^{2}}
+\Or(\dot{\phi},\ddot{\phi}/\dot{\phi})
\label{jarret0}\end{equation}
and 
\begin{equation}
\frac{G_{*}}{G}=1+\frac{\Omega_{1}^{2}(1-3\Omega_{m})^{2}}
{2\dot{\phi}^{2}+4\Omega_{2}(1-\Omega_{m})+3\Omega_{1}^{2}(1-2\Omega_{m})}
+\Or(\dot{\phi},\ddot{\phi}/\dot{\phi})\, .
\label{jarret}\end{equation}
 One could further substitute $\Omega_{2}$ from (\ref{eq:omega2})
but the resulting expressions are more cumbersome. The time variation
of the gravitational coupling to the same order is then equal to
\begin{equation}
\frac{\dot{G}_{*}}{G_{*}}=-6\frac{\Omega_{m}}{(3\Omega_{m}-1)^{2}}
\left(3\Omega_{m}+3\frac{G_{*}}{G}\Omega_{m}-2\right)
\left(1-\frac{G}{G_{*}}\right) \, .
\label{jarret2}\end{equation}
 In contrast to the small-$\alpha$ limit, this time the $\delta\phi$
perturbation does give significant contributions to $\gamma$ and
$G_{*}$. Furthermore the corrections to $\gamma$ and $G_{*}$ are
large even if we take $\dot{\phi}\rightarrow0$. This means that taking
a slow-roll solution is not sufficient to satisfy observational constraints.

As with the small-$\alpha$ limit, the contributions from $\xi_{3}$
and $\xi_{4}$ do not feature in the expressions for $\gamma$ and
$G_{*}$, although this time it is due to the smallness of $\dot{\phi}$
rather than $\delta\phi$. Finally, we notice that a chance cancellation
at first order of both non-Newtonian effects $\gamma-1$ and $\dot{G}_{*}/G$
(and $G_{*}/G-1$) in (\ref{jarret0}) and (\ref{jarret}) occurs for
$\Omega_{m}=1/3$, a remarkable value indeed. In this case it is
necessary to go to higher order in $\dot{\phi}$ .

\subsection{Slow-roll, small-$\alpha$ regime}

We now reduce ourselves to the simplest case, in which we take both
the slow-roll approximation and the small-$\alpha$ limit. Taking
$\ddot{\phi}\ll\dot{\phi}\ll1$, allows us to neglect all terms containing
$\xi_{2,3,4}$. The scalar field equation then implies that $U'$
is also small, and so in the limit we are considering the dark energy
must be mostly cosmological constant. Notice that to derive
(\ref{jarret0}) and (\ref{jarret})
we neglected terms of order $\alpha\dot{\phi}$ and, as a consequence,
the correction turned out to be of order $\alpha^{2}$. It is therefore
not correct to take the limit $\alpha\rightarrow0$ in (\ref{jarret0}),
(\ref{jarret}) and (\ref{eq:ac}), (\ref{eq:Gsa}) have to be used
instead. Equation~(\ref{eq:ac}) in the slow-rolling regime gives
\begin{equation}
\gamma\approx\frac{A}{C}=1-\Omega_{1}\left(1+\frac{3}{2}\Omega_{m}\right)+\Or(\alpha^{2},\dot{\phi}^{2},\ddot{\phi})\, .\label{eq:ac0}\end{equation}
Similarly from (\ref{eq:Gsa}) we obtain, after some simplification,
\begin{equation}
G_{*}=G\left(1+\frac{4+3\Omega_{m}}{2}\Omega_{1}\right)+\Or(\alpha^{2},\dot{\phi}^{2},\ddot{\phi})\, .\label{eq:slowg}\end{equation}
 Differentiating and simplifying this gives a particularly simple
relation \begin{equation}
\frac{\dot{G}_{*}}{G_{*}}=-\frac{21}{2}\Omega_{m}\Omega_{1}+\Or(\alpha^{2},\dot{\phi}^{2},\ddot{\phi})\, .\end{equation}
 Notice that both $\gamma$ and $\dot{G_*}/G_*$ are independent of the
potential. The slow-roll and small-$\alpha$ limit allows us to put
constraints directly on $\Omega_{1}$, regardless of the potential
$U$.

\section{Observing a coupled Gauss-Bonnet gravity}
\label{sec:obs}

We have seen that we can derive the two LPN observable quantities
$\gamma$ and $\dot{G_*}/G_*$ in terms of the GB parameters. Now we discuss
briefly how to use these quantities to put limits on the model. We
will explicitly describe two methods: growth of matter perturbations 
and the integrated Sachs-Wolfe (ISW) effect. In both cases we will see
that constraints from current data are rather weak, but non-trivial;
moreover, the prospects for future experiments are encouraging. For
simplicity, we consider only the limit of slow-rolling
and small $\alpha$ in this section.

\begin{figure}
\centerline{\includegraphics[bb=70bp 500bp 400bp 730bp,clip]{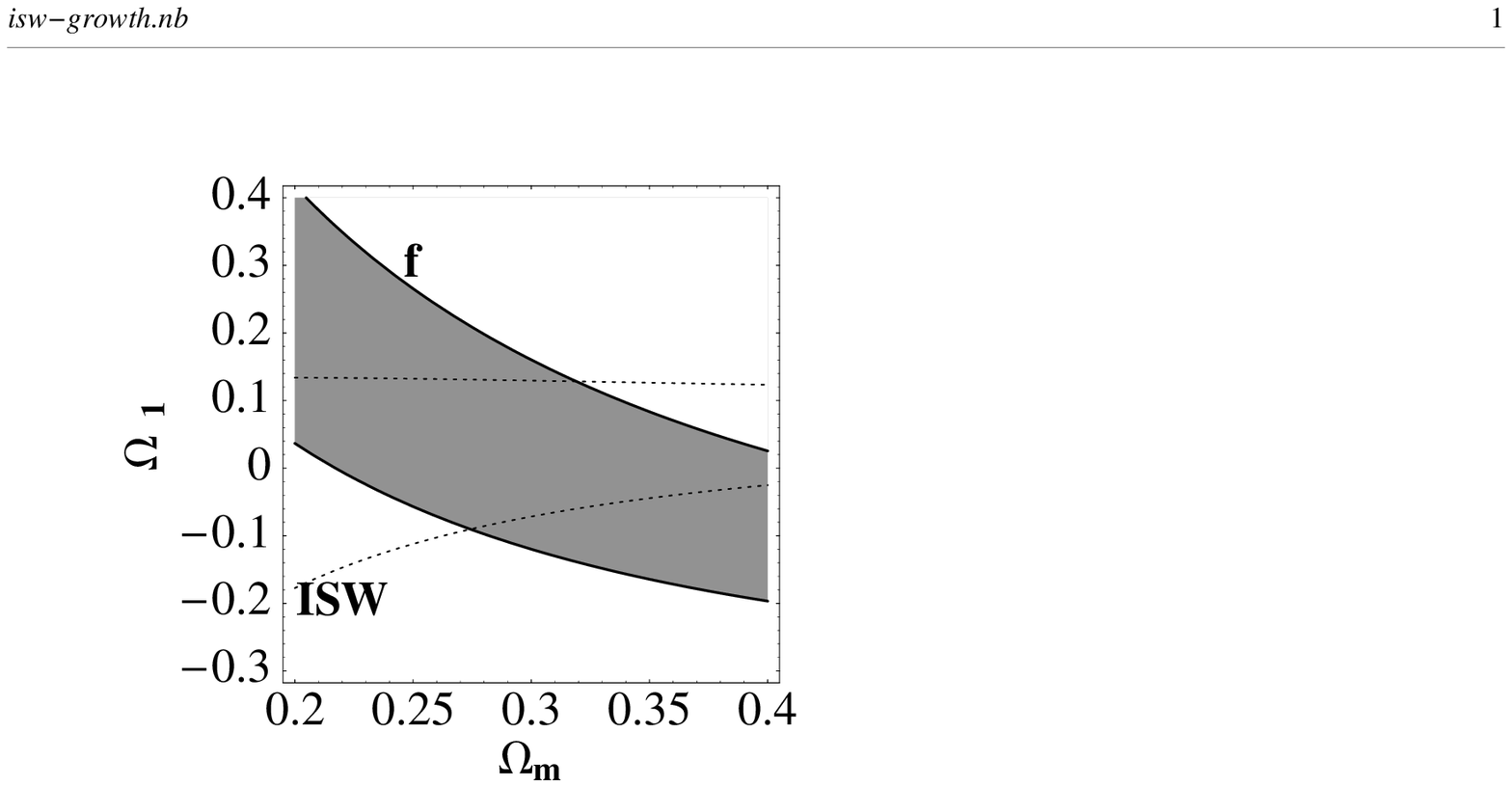}}
\caption{In this plot we show the constraints on $\Omega_{1}$ from the
  ISW effect (within the dotted lines) and from the growth rate $f$ of
  structures (grey region), assuming a present value $\Omega_{m}=0.3\pm0.1$.}
\label{iswfig}
\end{figure}

The equation for the matter growth that we get from (\ref{eq:slowg}) is
\begin{equation}
\ddot{\delta}+\left(1+\frac{\dot{\HH}}{\HH}\right)\dot{\delta}
-\frac{3}{2}\left(1+\frac{4+3\Omega_{m}}{2}\Omega_{1}\right)
\delta\Omega_{m}=0 \, .
\end{equation}
This is just the standard perturbation equation with an effective
matter parameter
\begin{equation}
\Omega_{{\rm eff}}\equiv
\left(1+\frac{4+3\Omega_{m}}{2}\Omega_{1}\right)\Omega_{m} \, .
\end{equation}
We can therefore obtain an approximate solution for the growth
factor $f\equiv\dot{\delta}/\delta$ in the standard way. In
conventional gravity this is $f=\Omega_{m}^{q}$
with the numerical coefficient $q\approx0.6$~\cite{jim}. To include
the effects of the GB term, we simply replace $\Omega_m$ with
$\Omega_{{\rm eff}}$. For small $\Omega_{1}$ this implies
\begin{equation}
\frac{\dot{\delta}}{\delta}=\Omega_{m}^{q}\left(1+q\frac{4+3\Omega_{m}}{2}
\Omega_{1}\right) \, .
\end{equation}  
For typical values of $\Omega_{m}=0.3$ and $q=0.6$, we see that
the GB term increases the standard $\Lambda$CDM growth rate by the
fraction $\approx1.5\Omega_{1}$. Comparing with the observational
result $f_{0}=0.51\pm0.1$~\cite{eisen} we obtain for the present
value $-0.2 \lesssim \Omega_{1}\lesssim 0.4$, as we show in
Figure~\ref{iswfig}. There are 
many upcoming observational programmes to detect the evolution
of clustering at various redshifts, so future data will no doubt
strengthen these constraints and extend their temporal range.

As a second observational constraint we consider the ISW effect, which depends
on the variation of Newton's potential $\Psi$. From (\ref{eq:Poisson})
we have for every wavenumber $k$\begin{equation}
\frac{\dot{\Psi}}{\Psi}=-1+\frac{\dot{\delta}}{\delta} + \frac{\dot{G_*}}{G_*}
\end{equation}
where we used the relation 
$\Omega_{m}\mathcal{H}^{2}\sim\rho_{m}a^{2}\sim a^{-1}$. Then we obtain
\begin{equation}
\frac{\dot{\Psi}}{\Psi}=-1+\Omega_{m}^{q}
\left[1+\frac{q(4+3\Omega_{m})}{2}\Omega_{1}\right]
-\frac{21}{2}\Omega_{m}\Omega_{1} \, .
\end{equation}
The standard $\Lambda$CDM result is then modified by a term that
for $\Omega_{m}\approx0.3$ amounts to $-2.4\Omega_{1}$. Comparing
with the observational result is rather difficult here, because
observational groups typically only quote their results in terms of
the ISW cross-correlation with large scale structure. Hence knowledge
would be required of the background dynamics, and of the full evolution
of perturbations at all times, which is beyond the scope of this work.
Moreover the uncertainties in the data so far are so large, even
for $\Lambda$CDM models, that a very detailed comparison looks premature.
A rough limit on $\Omega_{1}$ can be obtained taking the range
of $\Omega_{m}$ values that are consistent with ISW measurements
for a pure $\Lambda$CDM or for a constant-$w$ dark energy model.
 From~\cite{tommaso,gaz,coras} we can conservatively summarise the allowed
range as $0.05 \lesssim \Omega_{m} \lesssim 0.5$. We then see from
Figure~\ref{iswfig} that $\Omega_{1}$
has to be in the interval $(-0.2,0.1)$ in order not to produce too
large or too small a ISW cross-correlation. Apparently the constraints
from the ISW effect are similar to those from the structure growth; however,
let us remark that this procedure assumes implicitly that in our
coupled GB model the evolution of large scale structure is similar
to standard dark energy models, which is clearly unwarranted.

Combining the two constraints and assuming $\Omega_{m}=0.3\pm0.1$
at the present, we see that current data put a weak but not totally
irrelevant constraint on $\Omega_{1}$, of the order of
$|\Omega_{1}| \lesssim 0.15$.  This shows, firstly that the coupled GB term
is unlikely to provide the bulk of the dark energy, and secondly that
future data have the potential to greatly improve this bound. Finally,
the fact that $\Omega_{1}$  has to be quite a bit smaller than
unity is in agreement with our  assumptions of small $\alpha$ and
slow-rolling. Of course all of this says nothing about the value of
$\Omega_{1}$ in the past.

\section{Conclusion}

\label{sec:conc}

We have investigated extensions of quintessence models in which part
of the dark energy comes from quadratic-order gravity terms (and corresponding
scalar kinetic terms). Even though the models contained no direct
coupling of the quintessence scalar field to matter, the effective
gravitational coupling ($G_{*}$) in the model acquired a time dependence
and the stress parameter $\gamma$ deviates from its general-relativistic
value. We included all possible ghost-free second-order gravity terms
in our model, since it is natural for them to all be present simultaneously.
This can seen, for example, by considering a scalar field which arises
from the compactification of an $N$-torus.

The time variation of $G_{*}$ and the value of $\gamma$ depend on
the contribution of the higher-order gravity terms in the Friedmann equation,
i.e.\ on the density fractions $\Omega_{i}$. This suggests that
by taking them to be small, the constraints can be satisfied. If we
take the coupling of the higher-order gravity terms ($\alpha$) to be small,
this is indeed the case. On the other hand if we suppose that the
variation of the scalar field is small (which the $\Omega_{i}$
are all proportional to), then we find that $\gamma-1$ and $\dot{G}_{*}/G_{*}$
can still be large. This is due to the fact that the higher-order gravity
terms can still have a significant effect on perturbations to the
scalar field, even when they have little effect on the Friedmann equation.

We considered the possibility of observing the coupled GB term through
the matter perturbation growth and through the ISW effect. We found
that current data limit $|\Omega_{1}|\lesssim 0.15$, assuming both
small $\alpha$ and slow rolling. This implies that 
the present effect of a coupled GB term is quite limited; on the other hand,
its impact during other epochs in cosmic history remains unbounded.

The contributions from the higher-order {}``$k$-essence'' scalar kinetic
terms ($\Omega_{3,4}$) only affect $\gamma$ and $G_{*}$ indirectly
via the dynamics of the scalar field, and so the direct constraints
on them are much weaker (and completely absent in our slow-roll limit).
However, unless the couplings $\xi_{3,4}$ are unnaturally large,
the density fractions $\Omega_{3,4}$ will be smaller than $\Omega_{1}$
by factors of $\dot{\phi}^{2},\dot{\phi}^{3}$ respectively, and thus
completely negligible in a slowly rolling expansion. Moreover, we have
seen that the slow-roll condition is the less restrictive one, in
the sense that any faster dynamics will in general yield (barring
chance cancellation) tighter constraints on $\Omega_{i}$ (we
have shown this explicitly only in the small-$\alpha$ limit, but
we can conjecture it is a general property, since higher-order terms
in $\alpha$ will generally also bring in higher-order terms in the
time derivatives of $\phi$).
Of course it is possible that the corrections to $G_{*}$ coming from
the different modifications to gravity will cancel each other, resulting
in far weaker constraints. This will generally require fine-tuning,
although it could occur naturally as a result of symmetries in the
underlying theory. 

Finally, we emphasise that the constraints obtained here are restricted
to the Einstein frame and to energy conservation. This hypothesis
permitted us to disentangle any deviations from general relativity
exclusively from the higher-order terms. Relaxing this hypothesis, and also
the determination of exact cosmological solutions, are interesting
subjects that we leave for future study{\footnote{During the revision
of this paper, \cite{mota} appeared where similar Gauss-Bonnet
quintessence models were considered and compared with other
cosmological data.}}.

\ack It is a pleasure to thank Gilles-Esposito Farese
for helpful comments and discussions. CC thanks Martin Bucher and
Nemanja Kaloper for numerous interesting discussions. 
SCD acknowledges the Swiss Science Foundation and the Netherlands
Organisation for Scientific Research (NWO) for financial support.

\end{document}